\documentclass[12pt]{article}  
  
\def\12{\frac{1}{2}}
\def\14{\frac{1}{4}}

\def\tr{\mbox{tr}}

\def\a{\alpha}

\def\s{\sigma}  
\def\d{\delta}

\def\th{\theta}

\def\z{\zeta} 
\def\t{\tau}

\newcommand{\be}{\begin{equation}} % \label{#1#2#3}}     % non-hyper
\newcommand{\ee}{ \end{equation}}
\newcommand{\ba}{\begin{array}}
\newcommand{\ea}{\end{array}}
\newcommand{\bea}{\begin{eqnarray}}
\newcommand{\eea}{\end{eqnarray}}

\renewcommand{\a}{\alpha}

\renewcommand{\d}{\delta}
\newcommand{\pa}{\partial}

\def\tr{\mbox{tr}}

\begin{document}  
  
\begin{titlepage}

\begin{flushright}  
 
HUB-EP-00/34\\
   
\end{flushright}  
  
\vspace{3cm}  
%%%%%%%%%%%%%%%%%%%%%%%%%%%%%%%%%%%%%%%%%%%%%%%%%%%%%%  
\begin{center}  
  
{\Large \bf{Dynamics of Wilson Observables in Non-Commutative Gauge 
Theory}}  
  
\vspace{1cm}  
  
Mohab Abou-Zeid\footnote{abouzeid@physik.hu-berlin.de} and Harald 
Dorn\footnote{dorn@physik.hu-berlin.de} 

\vspace{.3cm}  
{\em Institut f\"{u}r Physik, Humboldt Universit\"{a}t zu Berlin, 
Invalidenstrasse 110, \\ D-10115   Berlin, Germany}   

\vspace{2cm}  
%%%%%%%%%%%%%%%%%%%%%%%%%%%%%%%%%%%%%%%%%%%%%%%%%%%%%%  
\begin{abstract}

An equation for the quantum average of the gauge invariant Wilson loop in
non-commutative Yang-Mills theory with gauge group $U(N)$ is obtained. In
the 't~Hooft limit, the equation reduces to the loop equation of ordinary
Yang-Mills theory. At finite $N$, the equation involves the quantum averages
of the additional gauge invariant observables of the non-commutative theory
associated with open contours in space-time. We also derive equations for 
correlators of several gauge invariant (open or closed) Wilson lines. 
Finally, we discuss a perturbative check of our results. 

\end{abstract}
  
\end{center}  
%%%%%%%%%%%%%%%%%%%%%%%%%%%%%%%%%%%%%%%%%%%%%%%%%%%%%%%  
\vspace{2cm}
 
\flushleft{September, 2000}
 
\end{titlepage}

\section{Introduction}
In the approach to gauge theory pioneered by Mandelstam~\cite{SM} and
Wilson~\cite{KW}, the basic 
dynamical object is the phase factor (or parallel transporter)
\be
U [C,A]  =  P \exp i \int_C A_\mu (x(\t) )  dx^\mu (\t ) .
\label{Wilson}
\ee
Here $P$ denotes the Dyson path ordering operation, the gauge field $A_\mu$  
is 
represented by matrices lying in the Lie 
algebra of the gauge group and the line integral runs over a closed loop $C$
parametrised by $x(\t )$, $0\le \t \le 1$. 
In the quantum theory, one would like to compute the gauge-field average of
the (gauge-invariant) trace of $U[C,A]$ weighted by the gauge field 
action\footnote{Below,
we will often omit indicating the dependence on $A$ for $U$ and $W$.} ,
\be
\langle W[C,A] \rangle  = \frac{1}{Z}\int [DA]  e^{-\frac{1}{g^2} 
\int F_{\mu\nu}F^{\mu\nu}} \frac{1}{N}\tr U[C,A] 
\ee 
($F_{\mu\nu}$ denotes the 
Yang-Mills field strength, and $Z$ is a normalisation factor). In the 
strong coupling limit, the loop average 
$\langle W[C]\rangle$ determines whether 
quarks are confined according to
 Wilson's area law~\cite{KW}. It also turns out to satisfy the so-called loop 
equations~\cite{AP1,MM1}, which are closed functional equations from which
the Feynman perturbation expansion can be shown to arise~\cite{MM2}. Thus 
quantum gauge theory can be reformulated in terms of such gauge-invariant 
Wilson observables. The loop equations in the 't~Hooft limit $N\to \infty$ with
$\lambda \equiv g^2 N$  kept fixed can be written as
\be 
\pa^\mu \frac{\d \langle W[C]\rangle}{\d \s^{\mu\nu} (x)} = 
-\lambda\int_C dy_\nu
\delta
 (x -y)  \langle W[C_{xy}]\rangle  \langle W[C_{yx}]\rangle  ,
\label{ordloop}
\ee
where $C_{yx}$ is defined as the part of $C$ from $x$ to $y$, $\pa^\a (x)$
is the path derivative at $x$ and $\d \s^{\mu\nu} (x)$ is the area derivative
in the $\mu , \nu$ plane at $x\in C$.  See e.~g.~\cite{AM,AP2} for a
detailed derivation and further explanation of these  equations.

In the present note, we will derive  equations analogous
 to~(\ref{ordloop}) for Wilson
observables in non-commutative gauge theory. The latter theory is a 
spatially 
non-local, higher derivative relative of ordinary gauge theory which exhibits
interesting perturbative behaviour~\cite{TF,MVN,MST}. Interestingly, 
the theory 
can be obtained in a 
zero-slope limit from open bosonic string theory in a constant background 
2-form gauge
potential~\cite{SW}. Using this connection to string theory, it was shown 
that the ordinary and non-commutative gauge theories are  classically 
equivalent~\cite{SW}, but it is not clear whether this persists at the 
quantum level. It 
is therefore important to reformulate the quantum non-commutative theory in
terms of gauge invariant observables satisfying geometric equations 
generalising~(\ref{ordloop}). This 
formulation, presumably, is suitable for the 
confining (strong coupling) phase of non-commutative gauge theory. 

Consider a $D$-dimensional Euclidean space with non-commutative coordinates, 
$[x^\mu , x^\nu ] = i\theta^{\mu\nu}$ (we assume the matrix $\th$ to be 
nondegenerate). The 
non-commutative Yang-Mills theory is obtained by working with commuting 
coordinates and replacing all ordinary products of gauge fields by
star products. The action for a $U(N)$ theory is
\be
S=\frac{1}{g^2}\int d x \tr \left( F_{\mu\nu}(x) \star F^{\mu\nu} (x) \right),
\label{NCaction}
\ee
where the non-commutative field strength is defined by\footnote{Henceforth,
 $A_\mu$ will denote the non-commutative gauge field which was denoted
by $\hat{A}_\mu$ in~\cite{SW}.}
\be
F_{\mu\nu} = \pa_\mu A_\nu -\pa_\nu A_\mu -i \left( A_\mu \star A_\nu -A_\nu 
\star
A_\mu \right)
\ee
and $\star$  is the star product defined by
\be
f(x) \star g(x) \equiv e^{\frac{i}{2}\theta^{\mu\nu} \frac{\pa}{\pa y^\mu}
 \frac{\pa}{\pa z^\nu}} f(x+y)g(x+z) |_{y=z=0} .
\ee 
The key properties of this product are associativity and cyclicity of the
integrals of $\star$-products over space. The cyclic property implies that 
the action~(\ref{NCaction})  is invariant 
under non-commutative gauge transformations. The classical field equations 
of the non-commutative gauge theory obtained by
varying~(\ref{NCaction}) are
\be 
D^\mu F_{\mu\nu} = 0 ,
\label{NCYMeq}
\ee
where $D_\mu = \pa_\mu -i [A_\mu , \cdot ]_\star $ is the covariant 
derivative. The Jacobi
identity
\be
\varepsilon^{\mu\nu\rho\sigma} D_\nu F_{\rho\sigma} = 0 \label{Bianchi}
\ee
holds whether the field equations are satisfied or not.

Consider an arbitrary (open or closed)  
contour $C$ in non-commutative space-time parametrised by 
$x+\xi (\t )$ with $0\le \t 
\le 1$. The generalisation of the Wilson 
factor~(\ref{Wilson}) in the non-commutative
theory is
\be
U [C,A] = P_\star \exp \left( i \int_C A_\mu (x+\xi (\tau))d\xi ^{\mu}(\tau )
 \right)
\label{NCwil}
\ee
where
$P_\star$ denotes path ordering along $x+\xi (\tau )$ from right to left
with respect to increasing $\tau $ of $\star$-products of functions. The star
multiplication is performed with respect to the variable $x$. The result
in (\ref{NCwil}) is independent of the splitting of the constant mode, i.e.
as long as $x+\xi (\tau )=x'+\xi '(\tau )$ one can replace $x$ and $\xi $
in (\ref{NCwil}) by $x'$ and $\xi '$.

Note that~(\ref{NCwil}) is not gauge invariant, even after taking the
trace over Lie algebra-valued matrices. If the contour 
$C$ is closed, this  contrasts with the situation in ordinary gauge 
theory, where a closed Wilson line is gauge invariant. For {\em closed} 
contours
the gauge invariant Wilson loop is defined by
\be
W_c[C]~=~\int dx \frac{1}{N}\tr U[x+C]~.
\label{3}
\ee
Here $x+C$ denotes the contour obtained by translating $C$ by $x$. In 
addition, it 
was found in~\cite{IIKK} that for an {\em open} contour $C$
the following open Wilson line with momentum $k_{\xi}$ 
\be
W_o[C]~=~\int dx \frac{1}{N}\tr U[x+C]\star e^{-ik_{\xi}x}
\label{4}
\ee
is gauge invariant provided the momentum $k_\xi$ and the distance 
$l =\xi (1)-\xi (0)$ 
between the end-points of $C$ satisfy the condition
\be
l^\nu =\theta^{\nu\mu}  k_\mu , \ \  k_{\xi}~=~\theta ^{-1}(\xi (1)-\xi (0)) .
\ee

In the quantum field theory,  the expectation value  of a given functional
$H[A]$ of the gauge field is understood in the sense of
\be
\langle H\rangle = \int [DA]~e^{-S[A]}~H[A]~,
\label{5}
\ee
with the normalisation $\langle 1\rangle =1$. Although on the classical
level and in a given gauge field configuration generically $U[x+C]\neq U[C]$,
the gauge field quantisation restores translation invariance:
\be
\langle U[x+C]\rangle ~=~\langle U[C]\rangle ~.
\label{5a}
\ee
Note that for any  $two$ $\star$-multiplied factors 
the order of factors under the $x$-integration can be interchanged. Moreover,
for factors satisfying suitable boundary conditions, the star multiplication
can be replaced by ordinary multiplication,
\be
\int d x f(x) \star g(x) = \int d x f(x) g(x) .
\label{2factors}
\ee
Using these facts, we find
\be
\langle W_o[C]\rangle ~=~(2\pi )^D \det \theta ~\delta (\xi (1)-\xi (0))
~\langle \frac{1}{N}\tr U[C]\rangle ~,
\label{7}
\ee
as well as ($V$ is the volume of $D$-dimensional space-time)
\be
\frac{1}{V}\langle \int dx \frac{1}{N}\tr U[C ]\rangle ~=~
\frac{\int dx}{V}\langle \frac{1}{N}\tr U[C ]\rangle~=~
\langle \frac{1}{N}\tr U[C ]\rangle ~.
\label{8}
\ee
Due to the $\delta $-function in (\ref{7}) it looks as if 
$\langle W_o[\xi ,A]\rangle $ would be nontrivial in the limit of closed 
contours only.
However, there are nontrivial correlation functions for more than
one contour \cite{DR,GHI}. Such correlation functions will indeed appear below.
Similarly, for the use in correlation functions it is crucial to keep
track of the $x$-integration for closed loops.
%%%%%%%%%%%%%%%%%%%%%%%%%%%%%%%%
\section{The loop equations}
We now turn to the derivation of the loop equations. The starting point
for this is the result
\be
\pa^\mu \frac{\d W [C] }{\d \s^{\mu\nu} (\xi (\t ))}= \frac{i}{N} 
\int dx \tr P_\star
\left( D^\mu F_{\mu\nu} (x+\xi (\t ) ) 
\exp \left( i\int_C A_\mu (x+\xi (\s ))d\xi^\mu {(\s)}
 \right) \right). 
\label{dMandel}
\ee
Since the derivation of this formula is based on purely geometrical
considerations together with the algebraic properties of the multiplication of
gauge fields, we can take it directly from the case of ordinary Yang-Mills;
see also \cite{AMNS2}.
%%%%%%%%%%%%%%%%%%%%%%%%

Thus the operator $\pa^\mu \d /
\d \s^{\mu\nu} (\xi )$ inserts the field equation~(\ref{NCYMeq}) at the 
point $x+\xi (\t)$. 
To compute the action of $\pa^\mu \d /
\d \s^{\mu\nu} (\xi)$ on the quantum average $\langle W_c [C] \rangle$, we
use the quantum field equation \footnote{For quantisation one has 
to add gauge fixing and 
ghost terms to 
the action. In the  case of ordinary Yang-Mills theory, it has been
shown in~\cite{GN} that the contribution of these terms to the field equation
inserted into the Wilson loops cancel. We assume this
to hold here, too.}
in a form which is well suited to keep track of subtleties in the evaluation
of functional derivatives of functionals $H$ built out of star products:
\be
\epsilon ~\Big <\frac{\delta S}{\delta A^a_{\nu }(Y)}~H\Big >~+~O(\epsilon ^2) ~=~
\langle H[A+\delta A]\rangle ~-~\langle H[A]\rangle ~,
\label{quantum}
\ee
with $A_{\nu }=A_{\nu }^aT_a~,~~\delta A^b_{\mu}(X)=\epsilon~\delta _{\mu\nu}\delta ^{ab}\delta (X-Y)$. In the following we will take\\
$H[A]=\frac{ig^2}{N}\tr P_\star \left( T_a (\t )\exp (i\int A_\mu (x+\xi )d
\xi^\mu )\right )$ and consequently
\be
\d A^b_\mu (x+\xi(\s)  ) = \epsilon ~\d_{\mu\nu} \d ^{ab}~\d (x+\xi 
(\s) -x' -\xi (\t) ) ~.
\label{special}
\ee
The notation $T_a (\t )$ indicates that the matrix $T_a$ representing one 
of the normalised generators of $U(N)$ is inserted at the parameter value $\t$.

Let us denote for a moment $X(\tau )=(x+\xi (\tau ))\in x+C$. The  
$\delta $-function
$\delta (X(\s)-X(\t))$ as a function of the argument
$X(\s )$ is a special case of a variation of the gauge field
as a function of $X(\s )$, as indicated 
in~(\ref{special}). On the other hand,  the star product
in~(\ref{dMandel}) refers
to the constant part in $X(\s)$. The 
second entry $X(\t)$
is a parameter which only determines where the $\delta $-function is peaked.
It is therefore $not$ involved in the evaluation of the star product.
This we indicate by our  notation. Only after evaluating the star product
one has to put $x'=x$. Making this identification from the very beginning
would factor the $\delta $-function out of the star product and lead to
incorrect results.

Note that
reparametrisation invariance and the cyclic property of the $\star$ product 
imply that
 we can choose the point $\xi (\t) $ where $\pa^\mu \d /
\d \s^{\mu\nu} (\xi )$ acts to be the endpoint $\z =\xi (1)$ of the 
contour. Then the
insertion of the field equation must be made at the  position mostly
to the left. Using~(\ref{2factors}), we can write
\be
\int dx \frac{\d S}{\d A^a_\nu (x+\z)} P_\star \left( T_a (1) e^{i\int 
A_\mu (x+\xi ) d\xi^\mu}\right) = \frac{1}{g^2} \int dx P_\star \left( 
D^\mu F_{\mu\nu} (x+\z) e^{i\int A_\mu (x+\xi )d\xi^\mu}\right) .
\ee
Thus we find\footnote{This equation was previously derived in~\cite{AMNS2},
however the key issue of  star product versus constant mode dependence was
not discussed there.} 
\bea
\frac{1}{V}\pa^\mu  \frac{\d}{\d \s^{\mu\nu} (\z )}
 \langle W_c [C] \rangle   & = &  - \frac{g^2}{NV}
\int d\xi_\nu (\t ) \Big< \int d x \tr P_\star \left( T_a (1 ) 
e^{i\int_\t^1
A_\mu (x+\xi (\s )) d\xi^\mu (\s )} \nonumber \right. \\ & & \left.
\d  (x+\xi (\tau )  -x' - \z  )
 T_a (\t ) e^{i\int_0^\t
A_\mu (x+\xi (\s )) d\xi^\mu (\s ) } \right) \Big|_{x=x'}\Big> .
~~~~~~~\label{inter1}
\eea  
The $U(N)$ 
colour group indices can be  rearranged using the factorisation
of the matrices $T_a$,
\be
(T_a )_{kl} (T_a )_{mn } = \d_{kn} \d_{lm} .
\label{Lie}
\ee
%%%%%%%%%%%%%%%%%%%%%

%%%%%%%%%%%%%%%%%%%%%%%
Then, applying formula~(\ref{delta}) of the appendix with $f$ and
$g$ represented by the Wilson factors in the integrand of~(\ref{inter1}) 
yields 
\bea
\frac{1}{V} \pa^\mu ~\frac{\d \langle W_c [C ] \rangle }{\d \s^{\mu\nu} (\z )}
& = &  - \frac{g^2}{NV}
\frac{1}{(2\pi )^D \det \th}\int d\xi_\nu (\t )  
\Big< \int d x \tr P_\star 
\left( e^{i\int_\t^1
A_\mu (x+\xi ) d\xi^\mu }\right) \star e^{-ik_{\xi (\tau )} x}   \nonumber \\ & & 
\cdot
\int d y \tr P_\star  \left( 
e^{i\int_0^\t
A_\mu (y+\xi ) d\xi^\mu }  \right)  \star e^{ik_{\xi (\tau )} y}   \Big> . 
\label{inter}
\eea   

Using the notation introduced above and reinstating differentiation at 
points  $\xi (\t)$ corresponding to arbitrary parameter values, (\ref{inter})
can be written as
\be
\frac{1}{V} \pa^\mu \frac{\d}{\d \s^{\mu\nu} (\xi )} \langle W_c[C] \rangle
 = - \frac{g^2N}{V} \frac{1}{(2\pi )^D \det \th} \int_C d\eta_\nu 
\langle W_o [C_{\xi \eta}] W_o [C_{\eta \xi }] \rangle .
\ee
This is the loop equation for the quantum average of the
gauge-invariant closed Wilson loop.  An interesting reformulation of this
equation is obtained by writing
\be
\langle W_o [C_{\xi \eta}] W_o [C_{\eta \xi}] \rangle = \langle 
W_o [C_{\xi \eta}] \rangle \langle  W_o [C_{\eta \xi}] \rangle
+\langle W_o [C_{\xi \eta}] W_o [C_{\eta \xi}] \rangle_{conn} 
\ee
(where  $\langle \ldots \rangle_{conn}$ denotes the connected part of a
correlation function) and applying eq.~(\ref{7}) to both $\langle 
W_o [C_{\xi \eta}] \rangle$ and $\langle  W_o [C_{\eta \xi}] \rangle$. The
product of delta functions which arises is dealt with in the usual way: 
writing
\be
[\d (\theta ^{-1}(\xi -\eta ) )]^2=\d (\theta ^{-1}(\xi -\eta ))\delta (0) =
\frac{V}{(2\pi )^D} \det\th
\d (\xi -\eta )~, 
\ee
we recover a delta function $\d (\xi -\eta )$ as in the ordinary loop
equation~(\ref{ordloop}) multiplied with a factor of $\det \th$. Thus the 
final 
result for the loop equation in the non-commutative theory at finite 
$N$ takes the form\footnote{The explicit 
powers of the (infinite) volume $V$ conspire 
to cancel trivial divergencies produced by overall translation
invariance of quantum averages.}
\bea
\frac{1}{V}\partial ^{\mu}\frac{\delta }{\delta\sigma ^{\mu\nu}(\xi )}
\langle W_c[C]\rangle & = & -\frac{\lambda}{ V^2}\int _C d\eta_{\nu}~
\delta (\xi -\eta )~\langle W_c[C_{\xi \eta }]\rangle~\langle 
W_c[C_{\eta \xi}]\rangle \nonumber \\ & & -\frac{g^2N}{(2\pi )^D 
V\det\theta}~
\int _C d\eta _{\nu}~\langle W_o[C_{\xi\eta}]~W_o[C_{\eta\xi}]
\rangle_{conn},~~~
\label{final}
\eea
for all points $\xi (\t) \in C$. 
In the 't~Hooft limit, the second term  on the r.~h.~s.\ involving
the connected part $\langle \ldots \rangle_{conn}$ 
of the two-point function vanishes and the equation looks 
like (\ref{ordloop}). Moreover for finite $N$ it is possible
to argue that, in the limit in which the non-commutativity parameter $\th$
is taken to zero,  the oscillatory behaviour
of the exponential factors in this connected part conspires to yield
a smooth limit in spite of the apparently divergent prefactor of
$1/\det\th$. Thus in this limit
 the equation is just the same as for standard
Yang-Mills theory. 

It is remarkable
that for finite $N$ the new gauge invariant objects for open contours
appear to be necessary for the description of the dynamics of closed
loops. In the non-commutative case for finite $N$ there is no overall  
$\delta $-function on the r.~h.~s.\ of the loop equation. 

In order to solve 
eq.~(\ref{final}) or eq.~(\ref{ordloop}), 
one should supplement it with the 
condition
\be
\varepsilon^{\mu\nu\rho\sigma} \partial_\nu 
\frac{\delta }{\delta\sigma ^{\rho\lambda}(\xi (\t) )}
\langle W_c[C] \rangle  = 0 ,
\ee
which follows from the Mandelstam formula~\cite{SM} and the Bianchi
identity~(\ref{Bianchi}).
%%%%%%%%%%%%%%%%%%%%%%%%%%%%%%%%%%%%%%%%%%%
\section{Loop equations for correlators}
Since the quantum average for a single observable $W_{o}$  
vanishes as a result of eq.~(\ref{7}), there
is no nontrivial quantum dynamics for these new object by themselves.
The situation changes if one considers correlation functions of several
closed and/or open contours. Here we consider as a first step the 
correlation function of
two Wilson loops. The generalisation to higher correlation functions 
and to the inclusion
of Wilson functionals for open contours is straightforward\footnote{For 
open contours, one has to be careful with possible subtleties
in the case where the variations hit one of the endpoints. There are
however no such problems if the difference vector $l^\mu$ is held fixed.}.
If one acts with the differentiation on loop $C^1$ the equation analogous
to~(\ref{inter1}) is ($\xi(1) \in C^1$)
\bea
\frac{1}{V}\partial ^{\mu}\frac{\delta }{\delta\sigma ^{\mu\nu}(\xi (1) )}
\langle W_c[C^1]W_c[C^2]\rangle
=  \frac{-g^2}{N^2V}\left (\int _{C^1}d\xi _{\nu}\langle \int dxdy 
\tr P_{\star} \{ T_a(1)T_a(\tau )\right .~~~~~\label{9}\\
\cdot ~U[C^1+x]\delta (x+\xi (\tau )-x'-\xi (1))\}
\vert _{x'=x}~\tr U[C^2+y]\rangle\nonumber\\  
\left .-\int _{C^2}d\eta _{\nu}\langle \int dxdy \tr \{T_aU[C^1+x]\}
\tr P_{\star}
\{U[C^2+y]T_a(\sigma )\delta (y+\eta (\sigma )-x-\xi (1)\}\rangle \right ). 
\nonumber
\eea
The first term looks like the r.h.s of (\ref{inter}) with $W_c[C^2]$
being a spectator only. Using the 
$\delta $-function to perform the $y$-integration and rearranging the
group indices using~(\ref{Lie}), the 
second term simplifies to
to
\bea
\mbox{term 2} ~= ~-\frac{g^2}{N^2V}\int d\eta _{\nu}(\sigma )\Big<
\int dx \left (P_{\star}e^{i\int _0^1A(x+\xi )d\xi }\right )_{nm}
~~~~~~~~~~~~~~~~~~~~~~~~~\label{10} \\
\cdot ~\left ( P_{\star}e^{i\int _0^{\sigma}A(x+\xi (1)-\eta (\sigma )+
\eta (\omega ))d\eta (\omega )}\right )_{mp}
~\star ~\left (P_{\star}e^{\int _{\sigma}^1A(x+\xi (1)-\eta (\sigma )+
\eta (\omega ))d\eta (\omega )}\right )_{pn}\Big> ~.
\nonumber
\eea 
A priori the first multiplication is no star product, but again due to the
$x$-integration we can replace it by a star multiplication. Then both
the star and matrix multiplications line up in such a way as 
to form the Wilson
factor for first going along a shifted version of $C^2$ 
(from $\eta$ back to $\eta$) and then along $C^1$ (from $\xi$ back to 
$\xi$), i.e.
\bea
\frac{1}{V}\partial ^{\mu}\frac{\delta }{\delta\sigma ^{\mu\nu}(\xi )}
\langle W_c[C^1]W_c[C^2]\rangle =~~~~~~~~~~~~~~~~~~~~~~~~~~~~~~~~~~~~~~~~~~~~~~~~~~~~ \label{11}\\
-\frac{g^2N}{(2\pi )^{D}V\det\theta}~
\int _{C^1}d\chi _{\nu}~\langle W_o[C^1_{\xi\chi}]~W_o[C^1_{\chi\xi}]~W_c[C^2]\rangle
\nonumber\\
-\frac{g^2}{NV}\int _{C^2}d\eta _{\nu}\langle ~W_c[C^1_{\xi}\circ (C^2_{\eta}+\xi -\eta )]~\rangle ~,
\nonumber
\eea
which is our equation for the correlation function of two Wilson loops.
Note that $W_c[C^1_{\xi}\circ (C^2_{\eta}+\xi -\eta )]$
can be written as
$$\int dx \frac{1}{N}\tr
\left ( U[x+C^1_{\xi}]\star e^{i\theta ^{-1}(\xi -\eta )}\star
U[x+C^2_{\eta}]\star e^{-i\theta ^{-1}(\xi -\eta )}\right )~.$$
%%%%%%%%%%%%%%%%%%%%%%%%%%%%%%%%%%
\section{Perturbative check}
Being very careful with the $\delta $-function under the star-multiplication
was the reason for obtaining a loop equation (before the $N\rightarrow\infty
$ limit) differing substantially 
from that in the usual Yang-Mills case for commutative space-time.
To illustrate this mechanism from a point of view slightly different from 
that above,
let us look at perturbation theory. For this purpose we consider a particular
diagram contributing in order $g^4$ to $\frac{1}{VN}\int dx\tr\langle
P_{\star}(D^{\mu}F_{\mu\nu}U)\rangle $. We will show that a $\delta $-function, 
forcing the two contours on the r.h.s. of the loop equation
to be closed, at finite $N$ appears in the commutative limit $\theta\rightarrow 0 $ {\it
only}. Considering all diagrams contributing to order $g^4$ is beyond the 
scope of this paper.

The insertion of the equation of motion at some point of the contour
gives a vertex which is the sum of contributions with one, two and three
gauge field legs. One has
\be
D^{\mu}F_{\mu\nu}~=~\partial ^2A_{\nu}-\partial _{\nu}\partial ^{\mu}A_{\mu}~+
~\mbox{terms quadratic and cubic in}~A~.
\label{a}
\ee
We consider 
\footnote{To simplify notation we take $N=1$ here.}
the insertion of the first summand at $\tau _4=1$ into the expansion of the
Wilson loop up to order $A^3$ (with the $\tau $-integration restricted by 
$0\leq\tau _1\leq\tau _2\leq\tau _3\leq\tau _4=1,~~\xi _i=\xi (\tau _i)$) 
\bea
I&=&\frac{i^3}{V}\int dx~\int 
\prod _{i=1}^3d\xi ^{\lambda _i}_i 
\langle \partial ^2A_\nu (x+\xi _4)\star A_{\lambda _3}(x+\xi _3)
...\star A_{\lambda _1}(x+\xi _1)\rangle _{g^4} \nonumber \\ 
&=&-\frac{i}{V}\int dx~\int 
\prod _{i=1}^3d\xi ^{\lambda _i}_i\exp (\frac{i}{2}
\sum _{1\leq i<j\leq 4}\partial _{\xi _i}\theta\partial _{\xi _j} )
\label{b} \\ & & \langle\partial ^2A_\nu (x+\xi  _4)A_{\lambda _3}(x+\xi _3)
...A_{\lambda _1}(x+\xi _1)\rangle _{g^4}\nonumber\\ 
&=&-ig^4\int 
\prod _{i=1}^3d\xi ^{\lambda _i}_i\exp (\frac{i}{2}
\sum _{1\leq i<j\leq 4}\partial _{\xi _i}\theta\partial _{\xi _j})
\{\big (\partial ^2 G_{\nu\lambda _2}(\xi _4-\xi_2 )G_{\lambda _3\lambda _1}
(\xi _3-\xi _1)\nonumber\\
&&~~~
+\partial ^2 G_{\nu\lambda _1}(\xi _4-\xi_1 )G_{\lambda _2\lambda _3}
(\xi _2-\xi _3)
+\partial ^2 G_{\nu\lambda _3}(\xi _4-\xi_3 )G_{\lambda _1\lambda _2}(
\xi _1-\xi _2)\}
~.
\nonumber 
\eea
We have denoted the gauge field propagator by $g^2G_{\mu\nu}$.
Choosing Feynman gauge we have $\partial ^2G_{\mu\nu}(x)=-
g_{\mu\nu}\delta (x)$. Denoting by $I_1$ the contribution from the 
first summand in the curly bracket of the last line of (\ref{b}), we find
\bea
I_1&=&ig^4g_{\nu\lambda _2}\int\prod _{i=1}^3d\xi ^{\lambda _i}_i\exp 
(\frac{i}{2}
\sum _{1\leq i<j\leq 4}\partial _{\xi _i}\theta\partial _{\xi _j})~
\delta (\xi _4-\xi _2)~G_{\lambda _3\lambda _1}(\xi _3-\xi _1)
\nonumber\\
&=&\frac{ig^4}{(2\pi )^{D/2}}~g_{\nu\lambda _2}\int\prod _{i=1}^3d
\xi^{\lambda _i}_i\int dk~
\delta (\xi _4-\xi _2+\theta k)~\tilde 
G_{\lambda _3\lambda _1}(k)~e^{ik(\xi _3-\xi _1)}
\label{c}
\eea
($\tilde G$ is the Fourier transform of $G$). 

For $\theta =0$ eq.~(\ref{c}) becomes
\be
I_1~=~ig^4g_{\nu\lambda _2}\int\prod _{i=1}^3d\xi ^{\lambda _i}_i
\delta (\xi _4-\xi _2) 
G_{\lambda _3\lambda _1}(\xi _3-\xi _1)~.
\label{d}
\ee
Then $I_1$ gets contributions only from points of the contour coinciding
with the point where the equation of motion is inserted by the contour 
variation.

For $\theta \neq 0$ eq.(\ref{c}) can be written as
\be
I_1~=~\frac{ig^4}{(2\pi )^{D/2}\det \theta}~g_{\nu\lambda _2}
\int\prod _{i=1}^3d\xi ^{\lambda _i}_i~\tilde G _{\lambda _3\lambda _1}
(\theta ^{-1}(\xi _2-\xi _4))~e^{i(\xi _3-\xi _1)\theta ^{-1}(\xi _2-\xi _4)}~.
\label{e}
\ee
Obviously, now all points of the contour contribute to $I_1$.\\

Since both eq.~(\ref{d}) (for $\theta =0$) and eq.~(\ref{e}) are simple 
reformulations of eq.~(\ref{c}) we see that the limit 
$\theta\rightarrow 0$ is smooth.

\vspace{.5cm}

Throughout this paper we have omitted questions of renormalisation. In 
ordinary Yang-Mills theory,   the renormalisation of Wilson loops is 
completely understood~\cite{GN,AMP,YA,DV,BNS}. On the other hand 
loop equations have 
been derived in a satisfactory manner in the presence of some regularisation 
only, although some remarks were made on the problems of such equations for 
renormalised Wilson 
loops~\cite{GN,YA,HD,PR}. In the present context, addressing these issues
will require a better understanding of the perturbative behaviour
found in~\cite{TF,MVN,MST}.\\[5mm] 
{\bf Acknowledgements}\\
We would like to thank Oleg Andreev for discussions and Yuri Makeenko for 
correspondence. M.\ A.\  is grateful to the Erwin Schr\"{o}dinger Institut  
and to Queen Mary and Westfield College for hospitality in Vienna and London 
during the early stages of this work. M.\ A.\ is supported by the Swiss 
National Science Foundation 
under grant 
number 83EU-056178. H.\ D.\ is partially supported by the 
Deutsche Forschungsgemeinschaft.

\section*{Appendix}

For any two functions $f,g$ with suitable boundary conditions, the identity
\be
\int dx f(x)\star \d (x+\xi (\t ) -x' -\z )\star g(x) |_{x=x'} = 
\frac{1}{(2\pi )^{D} \det \theta}
\int dx f(x)\star e^{-ik_\xi x} \int dy g(y) \star e^{ik_\xi y} \ \ ,
\label{delta}
\ee
where $k_{\xi (\t )} \equiv \th^{-1}
(\z -\xi (\t ) )$, is established as follows. Taking the 
star 
products on the l.~h.~s.\  yields
\be
\int dx \frac{dp_1}{(2\pi)^{D/2}} \frac{dp_2}{(2\pi)^{D}} 
\frac{dp_3}{(2\pi)^{D/2}} 
e^{-\frac{i}{2}(p_1 \th p_2 +p_2 \th p_3 +p_1 \th p_3 )} 
e^{i(p_1 + p_3 ) x} 
e^{ip_2 (x+\xi (\t ) -x'-\z )} \tilde{f}(p_1 ) \tilde{g}(p_3 ) .
\label{step1}
\ee
Setting $x=x'$ and integrating over $x$ imposes $p_1=-p_3$. Upon further 
integration 
over $p_3$ one finds 
\be
\int \frac{dp_1}{(2\pi)^{D/2}} 
\frac{dp_2}{(2\pi)^{D/2}} e^{-ip_1 \th p_2}
e^{ip_2 (\xi (\t ) -\z )} \tilde{f}(p_1 )\tilde{g}(-p_1 ) .
\ee
Integrating over $p_2$ imposes $\th p_1 = -(\xi (\t )-\z )$, so we are left 
with
\be
 \int  dp_1 \d (\theta p_1 + \xi (\t ) -\z ) 
\tilde{f}(p_1 )
\tilde{g}(-p_1 ) =  \frac{1}{\det \th} \tilde{f}
(\th^{-1} k_\xi ) \tilde{g} (-\th^{-1} k_\xi ) ,
\ee
which is identical to the r.~h.~s.\ of the desired result~(\ref{delta}).

\end{document}